# Scalar Supersymmetry

Alex Jourjine

FG CTP
Hofmann Str. 6-8
01277 Dresden
Germany

**Abstract**

We describe a new realization of supersymmetry, called scalar supersymmetry, acting in spaces of differential forms (bi-spinors), where transformation parameters are Lorentz scalars instead of spinors. The realization is related but is not reducible to the standard supersymmetry. Explicit construction of chiral multiplets that do not require doubling of the spectrum of a gauge theory is presented. A bi-spinor s-supersymmetric string action is described.

## 1. Introduction

It is well-known that the use of supersymmetry [1] to extend the Standard Model results in a number of attractive features of the models. Apart from curing the Higgs mass finetuning problem, supersymmetry leads to gauge coupling unification, provides candidates for Dark Matter, and sets the stage for gravity unification via superstrings. Unbroken supersymmetry requires that each observed particle has a superpartner with equal mass. Since the observed particle mass spectrum of the SM is not mass-degenerate, supersymmetry must be broken. Breaking supersymmetry is a non-trivial problem; it must be broken softly to preserve the desired cancellations of divergences, and presently there exist a number of phenomenologically viable supersymmetric extensions of the Standard Model [2, 3].

One characteristic feature of supersymmetry of such extensions appears in the particle spectrum even if supersymmetry is broken. Namely, each Standard Model particle must have a superpartner with spin differing by one-half. This is because in the Standard Model the bosonic gauge fields are real and transform in the adjoint representation of the gauge group $G_{SM} = SU(3)_C \times SU(2)_L \times U(1)_Y$ but the fermionic spinor fields are complex and transform in the fundamental representations of $G_{SM}$. As a result, one cannot combine the observed bosons and fermions into multiplets without violating gauge symmetry. In addition, the left and the right fermions couple differently to $SU(2)_L$. To accommodate the difference one is forced to use chiral supermultiplets. These can be only constructed if one pads each fermion with a superpartner of differing spin.

Despite an intensive search, most recently at LHC, no superpartners of the particle of the Standard model have been detected. One explanation for this failure could be that the predicted superpartners don't exist, which implies that a different realization of supersymmetry, the one that does not require doubling of the observed spectrum, must be used.

In this Letter we describe such a realization of supersymmetry. If implemented in a modification of the Standard Model, it would not require doubling of the Standard Model particle spectrum. We also show that the s-supersymmetry can be realized as symmetry of string action. With s-supersymmetry the observed gauge and fermion fields are allowed to mix through a supersymmetry transformation and no superpartners are needed. Parameters of s-supersymmetry are scalars instead of spinors, as is in the standard supersymmetry. It acts in spaces that are direct sums of spaces of commuting and anti-commuting differential forms and it requires the use of bi-spinor formalism [4, 5] to represent fermions. (Fermion bi-spinor fields are described by objects that transform as products of Dirac spinors and their Dirac conjugates.)

Although bi-spinors are seldom used for model building, the notion of bi-spinor is as old as that of Dirac spinor. In their anti-symmetric tensor form bi-spinors were discovered in 1928 by Ivanenko and Landau [6], in the same year Dirac proposed his theory of electron [7]. In fact, Ivanenko and Landau constructed an alternative to Dirac's solution of the electron's giromagnetic ratio problem[1]. However, the Ivanenko-Landau solution was more complicated than Dirac's by the standards of the time and naturally the latter won over as a basic descriptor of quantum fermionic matter.

Although bi-spinors have not been popular in phenomenology, they have been much in use in lattice gauge theory and, in particular, for building realizations of Dirac-Kähler twisting of the standard extended supersymmetry on the lattice [9, 10, 11, 12, 13]. Antisymmetric tensor form of bi-spinors also appears quite often in string theories in the form of $p$-forms, differential forms of fixed degree $p$. $P$-forms and their quantization have been studied both in supergravity and in string theory, including formulation of strings with two time parameters [14, 15, 16, 17, 18]. Theories of $p$-forms typically are restricted to commuting differential forms of a fixed degree. Here we will concentrate on the case where commuting and anticommuting inhomogeneous differential forms play equal role in the dynamics. For brevity we will concentrate on massless gauge fields and massless bi-spinors.

The Letter has three sections. In the following Section 2 we describe the needed basic ingredients of differential geometry. It can be skipped by readers familiar with the subject. Our results are contained in Section 3. Section 4 presents a brief summary. Appendix A contains an additional discussion that was deemed by one of the referees as too controversial for publication.

## 2. Differential Geometry, Z-basis, and Spinbeins

Although our results also apply when background gravity is present, to emphasize applications to phenomenology we will work with four-dimensional Minkowski space-time $M_4$ with metric $g_{\mu\nu} = diag(1,-1,-1,-1)$. All of the mathematical constructs we will use generalize with minor modifications to an arbitrary (pseudo-) Euclidean space-time. We will use the following index conventions: capital Latin letters $A, B, \ldots$ are reserved for the fermion generations, lower case Latin letters $a, b, \ldots$ are for gauge group representations, lower case Greek letters $\alpha, \beta, \ldots$ for spinor indices, while $\mu, \nu, \ldots$ for Lorentz tensor indices.

---

[1] Bi-spinors are also referred to as Ivanenko-Landau-Kähler (ILK) [8] or Dirac-Kähler (DK) spinors.

The basic notions of differential geometry that we need are the standard operations with differential forms on a manifold [19, 20], a basis in the space of differential forms, the $Z$-basis to define bi-spinors [21], and the spinbein decomposition of bi-spinors [22] to extract Dirac spinors from bi-spinors.

Given $M_4$ with coordinates $x^\mu$, a differential form $A$ in the coordinate basis (c-basis) is defined as a sum of homogeneous differential forms of degree $p$ with values in the Lie algebra of the gauge group $G$

$$A(x) = \sum_{p=0}^{4} A_p(x), \quad A_p(x) = A_{|\mu_1\cdots\mu_p|}(x) dx^{\mu_1} \wedge \cdots \wedge dx^{\mu_p}, \tag{1}$$

where $\wedge$ is the exterior product and $|\mu_1\cdots\mu_p|$ is a permutation of indices $\mu_1\cdots\mu_p$ with increasing order. In bi-spinor formalism such differential forms play the role of the fields of the standard (quantum) field theory.

Additional basic differential-geometric constructs that we need are the main automorphism $\alpha$, the main anti-automorphism $\beta$, and the contraction $(.,.)$ of a $p$-form $A_p$ with a $q$-form $B_q$ defined by

$$\alpha A_p = (-1)^p A_p, \qquad \beta A_p = (-1)^{p(p-1)/2} A_p,$$

$$(A_p, B_q) = \delta_{pq} tr(A_{\mu_1\cdots\mu_p})^\dagger B^{\mu_1\cdots\mu_p},$$

the exterior derivative $d$, $d^2 = 0$,

$$d : A_p \to A_{p+1}, \quad dA_4 = 0, \quad dA_p = \partial_\nu A_{|\mu_1\cdots\mu_p|} dx^\nu \wedge dx^{\mu_1} \wedge \cdots \wedge dx^{\mu_p}, \tag{2}$$

and the Hodge star operator $*$

$$* : A_p \to A_{4-p}, \quad (*A)_{4-p} = A^{|\mu_1\cdots\mu_p|} \varepsilon_{\mu_1\cdots\mu_p|\nu_1\cdots\nu_{4-p}|} dx^{\nu_1} \wedge \cdots \wedge dx^{\nu_{4-p}}, \tag{3}$$

$$** = (-1)^{p+1} = -\alpha, \tag{4}$$

where $\varepsilon^{\mu_1\cdots\mu_4}$ is the totally antisymmetric tensor of rank 4 with $\varepsilon^{0123} = 1$, $\varepsilon_{\mu_1\cdots\mu_4} = -\varepsilon^{\mu_1\cdots\mu_4}$. Very useful for us will also be operator $\star$, which we will call the chiral star operator, defined by

$$\star = -i * \alpha\beta = -i\beta *, \tag{5}$$

$$\star \star = 1. \tag{6}$$

From $d$ and $*$ the covariant divergence operator $\delta$ is defined by

$$\delta : A_p \to A_{p-1}, \quad \delta A_0 = 0, \tag{7}$$

$$\delta = *d*, \quad \delta^2 = 0. \tag{8}$$

We define a scalar product $\langle A, B \rangle$ of differential forms $A, B$ by linearity from

$$\langle A_p, B_q \rangle = \delta_{pq} \int tr[\alpha A_p^+ \wedge * B_q] = \delta_{pq} \int d^n x \, (\alpha A_p, B_q). \tag{9}$$

Note that $-\delta$ is the adjoint of $d$ with respect to scalar product (9) and, therefore, $(d - \delta)$ is self-adjoint. For Euclidean space-time definition (9) must be modified by removing automorphism $\alpha$.

We now introduce the $Z$-basis in the space of differential forms and establish the connection between antisymmetric tensors and bi-spinors. Given a set of Dirac $\gamma$-matrices, $\gamma^\mu = \{\gamma^\mu_{\alpha\beta}\}$, such that $\{\gamma^\mu, \gamma^\nu\} = 2g^{\mu\nu}$, the defining property of the $Z$-basis, $Z = \{Z_{\alpha\beta}\}$, is that operator $(d - \delta)$ takes the form of the Dirac operator [21]

$$(d - \delta) Z = Z (i \gamma^\mu \partial_\mu). \tag{10}$$

$Z$ is an $4 \times 4$ matrix of differential forms[2]. Any differential form $A$ can be represented in the $Z$-basis as

$$A = tr(Z \, \Psi(A)), \tag{11}$$

where $\Psi(A) = \{\Psi_{\alpha\beta}(A)\}$ are the coefficients of the representation and the trace is over the $\gamma$-matrix indices. Using (10) we obtain an explicit expression for $Z$ [21]

$$Z = \sum_p \gamma_{\mu_p} \cdots \gamma_{\mu_1} dx^{|\mu_1} \wedge \cdots \wedge dx^{\mu_p|}. \tag{12}$$

Since differential forms do not depend on the basis in which they are defined, the coefficients $A_{\mu_1 \cdots \mu_p}(A)$ of $A$ in the c-basis and the coefficients $\Psi_{\alpha\beta}(A)$ of $A$ in the $Z$-basis represent the same mathematical object. Also the transformation properties of the two sets of coefficients can be derived from basis independence of $A$: under Lorentz transformation $x \to \Lambda x$ the set $\{A_{\mu_1 \cdots \mu_p}(A)\}$ transforms as a collection of antisymmetric tensors, while $\Psi_{\alpha\beta}(A)$ transforms as

$$\Psi(A) \to S(\Lambda) \, \Psi(A) \, S(\Lambda)^{-1}, \tag{13}$$

where $S(\Lambda)$ is the spinor representation of the Lorentz group. Transformation (13) is the transformation law for bi-spinors: by definition they transform as a product of a Dirac spinor and its Dirac conjugate. Thus, we can identify the space of all $\Psi$ with the space of bi-spinors. Relations between the two sets of coefficients $\{A_{\mu_1 \cdots \mu_p}(A)\}$ and $\Psi_{\alpha\beta}(A)$ are derived using (12) and the completeness relations for $\gamma$-matrices

---

[2] For notational convenience our definition of $Z$ is the transposed of that in [21].

$$tr\left(\left[\gamma^{|\mu_1} \cdots \gamma^{\mu_p|}\right]\left[\gamma^{|\nu_1} \cdots \gamma^{\nu_q|}\right]^+\right) = 4\delta^{pq}\,\delta^{\mu_1\nu_1}\cdots\delta^{\mu_q\nu_p}\,, \tag{14}$$

$$\sum_p \left[\gamma^{|\mu_1} \cdots \gamma^{\mu_p|}\right]^*_{\alpha\beta}\left[\gamma^{|\mu_1} \cdots \gamma^{\mu_p|}\right]_{\gamma\delta} = 4\delta_{\alpha\gamma}\,\delta_{\beta\delta}\,. \tag{15}$$

It is given by

$$\left(A_p\right)_{|\mu_1\cdots\mu_p|}(\Psi) = tr\left(\gamma_{\mu_p}\cdots\gamma_{\mu_1}\Psi\right), \tag{16}$$

$$\Psi(A_p) = \frac{1}{4}\sum_p \gamma^{|\mu_1}\cdots\gamma^{\mu_p|}A_{|\mu_1\cdots\mu_p|}\,. \tag{17}$$

One property of $Z = \{Z_p\}$ that we will need below to define chirality of differential forms is

$$-i*\alpha\beta Z_p = Z_{n-p}\,\gamma^5,\ \gamma^5 = i\gamma^0\cdots\gamma^3\,. \tag{18}$$

Using the property we obtain for any differential form $A = tr(Z\Psi)$

$$-i*\alpha\beta A = tr\left(Z\,\gamma^5\Psi(A)\right). \tag{19}$$

We can now define chiral differential forms $A_{L,R}$ using projection operators $P_{L,R}$ constructed with the use of chiral star operator (5)

$$A_{L,R} = P_{L,R}A\,, \qquad P_{L,R} = \frac{1}{2}(1 \mp \star)\,, \qquad P_{L,R}^{\ 2} = 1\,, \qquad P_L P_R = 0\,. \tag{20}$$

Note that on $M_4$ chiral projection operators (20) can be defined only if $A$ is complex-valued. This can be seen from (5). However, this is sufficient for our purposes. The situation is different for Euclidean manifolds, where $(*\beta)^2 = 1$ and one can define real chiral differential forms [21]. From (18-20) we obtain that in the $Z$-basis the coefficients of the chiral differential forms are chiral bi-spinors

$$\left(1 \pm \gamma^5\right)\Psi(A_{L,R}) = 0. \tag{21}$$

Other useful commutator properties of the operators we introduced are

$$\alpha\beta = \beta\alpha\,, \qquad *\alpha = \alpha*\,, \qquad *\beta = \beta*\,, \qquad (d-\delta)\alpha = -\alpha(d-\delta)\,, \tag{22}$$

$$\alpha P_{L,R} = P_{L,R}\,\alpha\,, \qquad (d-\delta)P_{L,R} = P_{R,L}(d-\delta)\,. \tag{23}$$

The last ingredient we need is the spinbein decomposition of bi-spinors that extracts Dirac spinors from $\Psi$ transforming in some representation of the gauge group: $\Psi = \{\Psi^{ab}\}$. It is only needed to justify the form of the fermionic action. The extraction is done by using a spinbein $\eta^{aA}$ $a = 1,\ldots,N_\eta$, $A = 1,\ldots,4$, that is a multiplet of four commuting normalized Dirac spinors transforming in some $N_\eta$-dimensional representation of the gauge group $G$

$$\bar{\bar{\eta}}^{aA} = \Gamma^{AB}\bar{\eta}^{aB}, \qquad \Gamma^{AB} = diag(1,\ 1,-1,-1), \qquad \bar{\bar{\eta}}^{aA}_\alpha \eta^{aB}_\alpha = \delta^{AB}, \qquad (24)$$

where $\bar{\eta}$ denotes the Dirac conjugate of $\eta$. Spinbein decomposition of a bi-spinor is the ansatz [22]

$$\Psi^{ab} = \psi^{aA}\bar{\bar{\eta}}^{Ab}, \qquad (25)$$

where four generations of Dirac spinors $\psi^{aA}$, $a = 1,\ldots,N_\psi$, transform in a $N_\psi$-dimensional representation of $G$, which is not necessarily the same as that for the spinbein. Note that the form of spinbein decomposition (24, 25) implies that there are no right chiral bi-spinors: equation $\Psi(1\pm\gamma^5) = 0$ has no solutions.

The number of generations in (25) can be reduced from four to three or less if one uses a generally covariant constraint $\det\Psi^{ab} = 0$, where only Lorentz indices contribute to the determinant. The second known method to reduce the number of generations contained in a bi-spinor is the decomposition of $\Psi$ into minimal ideals of the associated Clifford algebra [5]. However, while coinciding with ours on $M_4$, this method is not generally covariant.

Given two general differential forms $F$, $H$, in the $Z$-basis we can write scalar product (9) as

$$\langle F,H\rangle = \int tr\left[\bar{\bar{\Psi}}(F)\Psi(H)\right], \qquad \bar{\bar{\Psi}}(F) = \gamma^0\,\Psi^+(F)\,\gamma^0. \qquad (26)$$

The appearance of $\gamma^0$ in (26) is the result of the presence of automorphism $\alpha$ in the definition of the scalar product (9). After spinbein anzatz (25) we obtain an equivalent representation of the scalar product in terms of Dirac spinor components

$$\langle F,H\rangle = \int tr\left[\bar{\bar{\psi}}^A(F)\psi^A(H)\right], \qquad \bar{\bar{\psi}}^A(F) = \Gamma^{AB}\bar{\psi}^B(F). \qquad (27)$$

In (26, 27) $\bar{\bar{\Psi}}(F)$ and $\bar{\bar{\psi}}^A(F)$ are bi-spinor conjugates of a bi-spinor and Dirac spinor, respectively.

### 3. Scalar Supersymmetry

To describe supersymmetry transformations we need to express the Lagrangian for gauge fields and fermions in terms of the basic operations defined in the previous section. In the $\xi$-gauge the Lagrangian for gauge fields, described by a connection $A_\mu = A^a_\mu\tau^a$, where $\tau^a, a = 1,\ldots,N_A$, are the generators of the Lie algebra of gauge group $G$, is given by

$$\mathcal{L}_g = -\frac{1}{2} tr(F_{\mu\nu} F^{\mu\nu}) + \frac{1}{\xi} tr(\partial_\mu A^\mu)^2, \quad A_\mu = A_\mu^a \tau_a, \quad tr(\tau_a \tau_b) = \frac{1}{2}\delta_{ab}, \tag{28}$$

where $F_{\mu\nu} = F_{\mu\nu}^a \tau_a$ is the curvature of the connection $A_\mu$ [3], the $\xi^{-1}$ term fixes the gauge, and $tr$ is the trace over the Lie algebra indices.

In terms of differential forms gauge fields are described by a commuting connection 1-form $A_1$, $A_1 = A_\mu dx^\mu$, while the curvature of the connection is given by 2-form $F$, $F = (1/2) F_{\mu\nu} dx^\mu \wedge dx^\nu$, $F = d_{A_1} A_1 \equiv (d + i g A_1 \wedge) A_1$, where $g$ is the coupling constant. Using the contraction of differential forms we can write the gauged-fixed Lagrangian for gauge fields as

$$\mathcal{L}_g = -\frac{1}{2} tr(dA_1 + ig A_1 \wedge A_1, dA_1 + ig A_1 \wedge A_1) + \frac{1}{\xi} tr(\delta A_1, \delta A_1), \tag{29}$$

where we used $\partial_\mu A^\mu = -\delta A_1$. The quadratic part of this Lagrangian that describes free fields is then given by

$$\mathcal{L}_g^0 = -\frac{1}{2} tr((d - \delta) A_1, (d - \delta) A_1) + \lambda\, tr(\delta A_1, \delta A_1), \quad \lambda = \left(\frac{1}{\xi} - \frac{1}{2}\right), \tag{30}$$

where $d^2 = \delta^2 = 0$ was used and, for convenience, we combined $d$ and $\delta$ in the first term.

We will now consider the fermionic fields. We will describe them by anti-commuting inhomogeneous differential forms $\Phi$ with values in the Lie algebra of the gauge group [5, 8, 21, 22]. The Lagrangian for $\Phi$ must be of the first order and, therefore, has the unique form given by

$$\mathcal{L}_f = tr(\alpha \Phi, (d_A - \delta_A) \Phi), \tag{31}$$

where $-\delta_A$ is the adjoint of $d_A$ with respect to (9). The free-field part of (31) is

$$\mathcal{L}_f^0 = tr(\alpha \Phi, (d - \delta) \Phi). \tag{32}$$

Comparing (29) with (31) we see that the single principle difference between gauge fields and bi-spinor fermions is their commutativity property. Otherwise, both are described by the same mathematical object. Notably, as mathematical constructs, Dirac spinors are quite different from gauge fields. They cannot even be defined on some space-times that are otherwise physically perfectly acceptable.

Using spinbein decomposition (25) with constant spinbein $\eta$ we obtain that in terms of physical Dirac spinor components $\psi^{aA}$ the Lagrangian (32) becomes

$$\mathcal{L}_f^0 = tr\, \overline{\overline{\Psi}} (i\partial) \Psi = tr\, \overline{\psi}^A (i\partial) \psi^A, \tag{33}$$

---
[3] We omit the ghost terms, since they are not relevant for our discussion.

where $\overline{\overline{\Psi}}, \overline{\overline{\psi}}$ are the conjugations of $\Psi, \psi$ defined in (26, 27). The reduction of (32) to (33) provides justification for the choice of the fermionic action (31).

Observe that Lagrangian (33) is an alternating sum of Lagrangians for four Dirac spinors $\psi^A$, two of which, those with $A = 1, 2$, enter the sum with the plus sign, while spinors with $A = 3, 4$ enter with the minus sign. The minus sign in the latter two terms has non-trivial consequences for quantization. Strictly speaking, the $A = 3, 4$ spinors are Dirac spinors only algebraically. Dynamically they are not Dirac spinors but rather anti-Dirac spinors: their action is the negative of Dirac spinor action and, hence, under the canonical quantization the assignment of creation and annihilation operators has to be reversed as compared to the standard Dirac spinor assignment. This is the only way one can ensure non-negativity of contribution of $A = 3, 4$ spinors to the quantum Hamiltonian of the system [22].

We will now describe a realization of supersymmetry in the space that is a direct sum of spaces of commuting and anti-commuting differential forms. Because the transformation parameters are Lorentz scalars, we shall call it scalar supersymmetry (s-supersymmetry). As we will see, gauge interactions always break s-supersymmetry. In an unbroken form it can only be realized for free field action in the particular $\xi$-gauge with $\xi = 2$.

We begin with the simplest case with $G = U(1)$ and zero mass. The combined $U(1)$ action for $\xi = 2$ gauge reduces to

$$S = -\frac{1}{2}\int d^4x \left(\alpha(d-\delta)A,(d-\delta)A\right) + \int d^4x \left(\alpha\Phi,(d-\delta)\Phi\right). \tag{34}$$

In (34) $A, \Phi$ is an arbitrary complex commuting or anticommuting differential form. To match the bosonic and fermionic degrees of freedom we were forced to promote $A_1$ in (29, 30) from a real commuting 1-form to an arbitrary commuting complex differential form $A$. We will call $(A, \Phi)$ a complex supermultiplet. With such a modification action (34) is invariant with regard to the transformation

$$\delta A = \varepsilon \Phi, \tag{35}$$

$$\delta \Phi = \frac{1}{2}\varepsilon^* (d-\delta)A, \tag{36}$$

where $\varepsilon$ is a complex-valued anticommuting transformation parameter. It is a Lorentz scalar. To derive invariance of (34) we used that $(d-\delta)$ is self-adjoint and that $\Phi$ and $\varepsilon$ anticommute. The former uses the Stokes theorem: $\int_{M_4} df = \int_{\partial M_4} f = 0$ for vanishing field contributions at infinity. Obviously, transformation (35) is a supersymmetry transformation: it mixes the bosonic and the fermionic degrees of freedom.

Representing (35, 36) in the $Z$-basis and using spinbein decomposition of $\Psi(A), \Psi(\Phi)$ we observe that (35, 36) do not mix generations of $\psi^{aA}$. This implies that the constraints $\det\Psi^{ab}(A) = 0$, $\det\Psi^{ab}(\Phi) = 0$ are consistent with scalar supersymmetry and (35, 36) are also symmetry transformations for bi-spinors containing three generations of (anti)-Dirac spinors.

It is easy to see that the commutator of two transformations in (35) is given by

$$[\delta_1, \delta_2] = \frac{1}{2}(\varepsilon_2^* \varepsilon_1 - \varepsilon_1^* \varepsilon_2)(d - \delta), \qquad (37)$$

while the anticommutator of the corresponding s-supersymmetry charges is given by

$$\{Q, Q^*\} = \frac{1}{2}(d - \delta). \qquad (38)$$

Expression (37) should be compared with the commutator of two transformations of the standard supersymmetry on $M_4$

$$[\tilde{\delta}_1, \tilde{\delta}_2] = 2\bar{\theta}_1 \gamma^\mu P_\mu \theta_2, \qquad (39)$$

where $P_\mu = i\partial_\mu$ is the translation operator and $\theta_k$, $k = 1, 2$, are infinitesimal Grassmann parameters transforming as Dirac spinors. We observe that for complex multiplet the standard and s-supersymmetry (35, 36) are related via the transformation of the bases in the space of differential forms $\{dx^{|\mu_1} \wedge \cdots \wedge dx^{\mu_p|}\} \to Z$ that maps $(d - \delta)\Phi$ into $i\hat{\partial}\Psi(\Phi)$.

The requirement that $A$ is complex-valued may be physically unacceptable. Hence, the simplest realization of s-supersymmetry most likely is an illustrative algebraic exercise. Note that, as can be seen from (5), unlike in Euclidean space-time, in Minkowski space-time there are no real bi-spinors. Therefore, to provide physically acceptable realizations of s-supersymmetry we have to restrict ourselves to real-valued gauge field differential forms but we cannot use real-valued fermionic forms. This means that to match the degrees of freedom we need to reduce their number for complex-valued fermions by half. The simplest way to do this is to use chiral fermionic differential forms we described in the preceding section[4]. In addition we have to use two left conversion operators: one that transforms real forms into left chiral complex forms and one that acts in the opposite direction. The most obvious left conversion operators are parameterized by a real parameter $\mu \neq 0$

$$K_A : A \to \Phi_L, \qquad K_A = \sqrt{2} P_L (1 + i\mu^{-1}(d - \delta)) P_+, \qquad (40)$$

$$K_\Phi : \Phi_L \to A, \qquad K_\Phi = \sqrt{2} P_+ (1 - i\mu^{-1}(d - \delta)) P_L, \qquad (41)$$

where the left chiral differential forms $\Phi_L$ are defined in (20) and

$$P_+ \Phi \equiv \mathrm{Re}\,\Phi = (1/2)(\Phi + \Phi^+) = (1/2)(1 + C)\Phi, \qquad C\Phi = \Phi^+, \qquad (42)$$

$$P_- \Phi \equiv \mathrm{Im}\,\Phi = (1/2i)(\Phi - \Phi^+) = (1/2i)(1 - C)\Phi, \qquad (43)$$

---

[4] Another way to cut the fermionic degrees of freedom in half is to use (anti)-Majorana spinors.

are projectors on the real and imaginary parts of a complex-valued differential form $\Phi$. Parameter $\mu$ of dimension of mass is needed in (40, 41) to compensate for the dimension of $(d-\delta)$. The right conversion operators are obtained from (40, 41) by $P_L \to P_R$.

Using (22, 23, 40-43) we obtain the most important properties of the left conversion operators

$$K_A K_\Phi : \Phi_L \to \Phi_L, \quad K_A K_\Phi = \left(1 + \mu^{-2}(d-\delta)^2\right)P_L, \tag{44}$$

$$K_\Phi K_A : A \to A, \quad K_\Phi K_A = \left(1 + \mu^{-2}(d-\delta)^2\right)P_+, \tag{45}$$

$$\alpha(d-\delta)K_\Phi = K_A^+ \alpha(d-\delta). \tag{46}$$

We can now describe s-supersymmetry realization for $U(1)$ $\xi = 2$ action for real Abelian massless bosonic fields $A_R$ and massless chiral bi-spinors with the action

$$S = -\frac{1}{2}\int d^4 x \left(\alpha(d-\delta)A_R, (d-\delta)A_R\right) + \int d^4 x \left(\alpha \Phi_L, (d-\delta)\Phi_L\right). \tag{47}$$

We will call pair $(A_R, \Phi_L)$ a chiral supermultiplet. Using projection operators (20, 42-43) we can rewrite it in an equivalent form

$$S = -\frac{1}{2}\int d^4 x \left(\alpha(d-\delta)P_+ A, (d-\delta)P_+ A\right) + \int d^4 x \left(\alpha P_L \Phi, (d-\delta)P_L \Phi\right). \tag{48}$$

In (48) $A$ is an arbitrary complex commuting differential form, while $\Phi$ is an arbitrary complex anticommuting differential form. Commutativity property of $A$, $\Phi$ is now the only property that distinguishes bosons from fermions. Action (47, 48) is invariant with regard to the infinitesimal transformation

$$\delta A = i\varepsilon K_\Phi \Phi, \qquad \delta \Phi = -\frac{1}{2}i\varepsilon K_A (d-\delta)A, \tag{49}$$

where, $\varepsilon$ is a real anticommuting transformation parameter and in addition to properties that were used to derive invariance of (34) under (35), we used (46). On-shell (49) reduces to (35, 36) with real $\varepsilon$.

We will now consider non-Abelian case with $G = SU(N) \times U(1)$ as an example. Extension to $G = SU(N_1) \times SU(N_2) \times U(1)$ of the Standard Model is straightforward. As we will presently see, the $U(1)$ factor in $G$ is actually a consequence of s-supersymmetry, needed to equalize the number of the bosonic and the fermionic degrees of freedom.

To match the degrees of freedom, in addition to promoting gauge field 1-form $A_1$ to an arbitrary real inhomogeneous real differential form $A$, we have to assign $A$ and $\Phi_L$ to appropriate representations of $G$. We have to keep in mind that $\Phi_L$ represents fermions and is an arbitrary chiral complex inhomogeneous differential form. Further, the physical (anti-)Dirac components of $\Psi = \Psi(\Phi_L)$, must transform in the fundamental representation of $G$, while the real gauge form $A$ must transform in the adjoint representations of the factors of $G$.

From these requirements we obtain that the simplest choice with equal number of degrees of freedom for $A$ and $\Phi_L$ is when $\Phi_L$ transforms in $N \times \bar{N}$, the direct product of fundamental and anti-fundamental representations of $G$. This representation is obtained if we use the spinbein decomposition of $\Phi_L$ with (anti-)Dirac fields $\psi_L^{aA}$ and spinbein $\eta^{aA}$ given by

$$\Phi_L^{ab} = tr(Z \Psi_L^{ab}), \quad \Psi_L^{ab} = \psi_L^{aA} \bar{\eta}^{bA}, \quad \bar{\eta}^{aA} = \Gamma^{AB} \bar{\eta}^{aB}, \quad \psi_{L,R} = \frac{1}{2}(1 \mp \gamma^5)\psi, \quad (50)$$

with $\psi_L^{aA}$, $\eta^{aA}$ transforming in the $N$ of $G$. At the same time $A$, which also has to transform in $N \times \bar{N}$ of $G$, separates into its irreducible components according to

$$A^{ab} = \frac{1}{N} B \delta^{ab} + W^{ab}, \quad B = tr A, \quad W^{aa} = 0, \quad a,b = 1,\ldots,N, \quad (51)$$

where $B$ transforms in the trivial, and $W^{ab}$ in the $(N^2 - 1)$-dimensional adjoint representations of $G$. Note that since $\eta^a$ are physical objects that are not observable as fields [22] our representation assignment matches the physical degrees of freedom but does not match the observable degrees of freedom. In fact, in our massless example the number of the observable gauge degrees of freedom per helicity state is $16N^2$ for bosons, while for fermions it is $4N$.

We can now write down s-supersymmetry transformations for left chiral scalar supermultiplet with free action in $\xi = 2$ gauge (the right chiral case is completely analogous)

$$S_0 = -\frac{1}{2}\int d^4x \, tr(\alpha (d-\delta)A, (d-\delta)A) + \int d^4x \, tr(\alpha \Phi_L, (d-\delta)\Phi_L), \quad (52)$$

where $A \equiv A_R$ and in terms of irreducible gauge field components $B$, $W$, of $A$ the gauge part of (52) is given by

$$S_g^0 = -\frac{1}{4}\int d^4x (\alpha (d-\delta)B, (d-\delta)B) - \frac{1}{2}\int d^4x \, tr(\alpha (d-\delta)W, (d-\delta)W). \quad (53)$$

Following the same steps as for (48, 49) we obtain that (52) is invariant under

$$\delta A = i\varepsilon K_\Phi \Phi, \qquad \delta \Phi = -\frac{1}{2} i\varepsilon K_A (d-\delta)A, \quad (54)$$

or, equivalently, under

$$\delta B = i\varepsilon\, K_\Phi\, tr\,\Phi, \qquad \delta W^{pq} = i\varepsilon\, K_\Phi \left( \Phi^{pq} - \frac{1}{N}\delta^{pq} tr\,\Phi \right),$$

$$\delta \Phi = -\frac{1}{2} i\varepsilon\, K_A \left( (d-\delta)\left( \frac{1}{N}\delta^{pq} B + W^{pq} \right) \right), \tag{55}$$

where $\varepsilon$ is an infinitesimal real Grassmann parameter.

Note that, because gauge fields are real, our realization of s-supersymmetry requires that the left- and the right-handed fermions couple to their own sets of gauge fields. Since experimentally we observe only one set of gauge fields that couple to left-handed fermions only, it follows from s-supersymmetry that the right-handed fermions have nothing to couple to and must be $SU(2)$ singlets.

We will now discuss what happens when we turn gauge interactions on. The simplest way to introduce gauge interactions is to use minimal gauging. In our case the minimally gauged Lagrangian for interacting fields is

$$\mathcal{L} = -\frac{1}{4}(dB,dB) - \frac{1}{2} tr\left(d_{W_1}W, d_{W_1}W\right) + tr\left(\Phi, (d_{A_1} - \delta_{A_1})\Phi\right), \tag{56}$$

$$d_{A_1} = d + i\, gB_1 \wedge + i\, g'W_1 \wedge, \tag{57}$$

where $A_1, B_1, W_1$ are the 1-form components of the expansions of $A, B, W$, and $g, g'$ are coupling constants for the $U(1)$ and $SU(N)$ factors of $G$. The Lagrangian (56) is invariant with respect to gauge transformations

$$B \to B + d\varphi, \tag{58}$$

$$W_1 \to \Omega(x) W_1 \Omega^{-1}(x) + \Omega(x) d\Omega^{-1}(x), \qquad \Omega(x) \in SU(N), \tag{59}$$

$$W_p \to \Omega(x) W_p, \quad p \neq 1, \tag{60}$$

$$\Phi \to \exp(i\phi_0)\Omega(x), \tag{61}$$

where $\varphi = \varphi(x)$ is an arbitrary real inhomogeneous differential form. It follows from Coleman-Mandula theorem [24] that (58-61) are the most general local symmetry transformations that can be imposed on a system where $A$ is an inhomogeneous differential form.

Supersymmetry transformations (54, 55) mix components $B_p(W_p)$ of gauge field $B(W)$ together and thus violate the special role $B_1(W_1)$ play in (56-61). We conclude that local gauge symmetry (58-61) breaks s-supersymmetry (54, 55) of the free part of the Lagrangian. It is an open question, whether linear realization of s-supersymmetry described here can be at most symmetry of free part of the Lagrangian or an extension to interacting Lagrangian exists. In any case some realization of s-supersymmetry must exist. This follows from the existence of conserved s-supersymmetric current that is a gauged version of s-supersymmetric current for free-field s-supersymmetry. For details we refer the reader to [25], where also bi-spinor BRST is described.

As a final remark we note that exact s-supersymmetry has a realization as a global supersymmetry of a string action. The action is a supersymmetric version of the bi-spinor string action described in [26]. Consider a collection of complex commuting and anticommuting 2-forms $B^A$ and $F^A$, $A = 0,\ldots,D-1$, transforming in some representation of a gauge group and defined on a two dimensional manifold with metric $g_{\mu\nu}$, $\mu,\nu = 0,1$ that is imbedded into $D$-dimensional Minkowski space-time $M_D$ with metric $\eta_{AB}$. Assume that $B^A$ and $F^A$ transform in the same representation of a gauge group. Then the action

$$S = \int \sqrt{-g}\,d^2 x\, \eta_{AB} tr\big((\alpha(d-\delta)B^A, (d-\delta)B^B) + (\alpha F^A, (d-\delta)F^B)\big). \tag{62}$$

is globally both gauge invariant and supersymmetric under the transformation

$$\delta B^A = \varepsilon F^A, \tag{63}$$
$$\delta F^A = -\varepsilon^* (d-\delta) B^A, \tag{64}$$

where trace is over the gauge group representation indices. Expanding

$$B^A = B_0^A + B_\mu^A dx^\mu + (1/2) B_2^A \varepsilon_{\mu\nu} dx^\mu \wedge dx^\nu, \tag{65}$$

and taking into account that

$$dB_0^A = \partial_\mu B_0^A dx^\mu, \quad \delta B_0^A = 0, \quad (\alpha\, dx^\mu, dx^\nu) = -g^{\mu\nu}, \tag{66}$$

we find that (62) contains two bosonic strings described by $\operatorname{Re} B_0^A$, $\operatorname{Im} B_0^A$. In the alternative, one can use left or right chiral differential forms for fermions and real differential forms for bosons. Then only one bosonic string described by real $B_0^A$ is contained in (62). How action (62) fits into the standard superstring classification and how its critical dimension depends on $D$ are open questions.

## 4. Summary

In summary, we presented a new realization of supersymmetry acting in the space of commuting and anticommuting differential forms. It could relieve supersymmetric models beyond the SM from requiring that each observed particle must have a superpartner particle. S-supersymmetry can only be possible if fermionic matter is represented by bi-spinors, instead of Dirac/Weyl spinors. S-supersymmetry with non-Abelian gauge fields requires the appearance of $U(1)$ factor in its gauge group.

We described explicit s-supersymmetry transformations for non-interacting bi-spinor gauge theory for complex and chiral multiplets. S-supersymmetry for complex multiplets can be reduced to the standard supersymmetry on space-times with spin structure. Chiral multiplet realization of s-supersymmetry seems to be genuinely different from the standard supersymmetry. The exact nature of interrelation needs more clarification. In any case, s-supersymmetry cannot be reduced to the standard supersymmetry on space-time where spinors cannot be defined. At least in this sense s-supersymmetry presents a novel type of transformations that mix bosonic and fermionic degrees of freedom.

Since the main benefit of supersymmetry is a cure for Higgs problem, it is not clear whether interacting s-supersymmetry would be needed for phenomenological models beyond the SM. After all, all that is needed is that the divergent contributions from fermion loops for Higgs self-energy cancel the bosonic ones. For the renormalized remainder supersymmetry may very well be broken from the beginning. Whether free-field s-supersymmetry indeed provides the Higgs mass problem cure is an open question.

Although we concentrated on explicit realization of free-field s-supersymmetry, the interacting bi-spinor gauge theory s-supersymmetry should exist. This follows from the existence of conserved s-supersymmetric current in interacting theory [25], which turns out to be the minimally gauged version of the current of free-field s-supersymmetry.

S-supersymmetry of interacting theory can also be realized in a superstring action, possibly providing an alternative way for construction of the theory of quantum gravity interacting with gauge fields and bi-spinor fermionic matter. String s-supersymmetry realization could also provide another, admittedly more circuitous, route to s-supersymmetry of interacting bi-spinor fields via the standard stacking of D-branes procedure.

Although bi-spinor gauge theory with SM gauge group is renormalizable by power count, the construction of full perturbative quantum bi-spinor modification of the SM has yet to be completed. However, some of its unusual features can be gleaned from its tree-level version, which can be easily constructed by the minimal gauging [22, 28]. One distinguishing feature that appears is that it admits explicit dimension three mass terms that are severely restricted in form [28]. This, in turn, leads to essentially unique forms of textures of the CKM and PMNS mixing matrices that agree with the experiment [27]. In addition it can predict the experimentally observed equality of two CKM matrix elements: $V_{ts} = V_{cb}$, something that the SM in principle cannot do. In distinction from all recent extensions of the SM, the observed textures appear without addition of new degrees of freedom.

It is Dirac spinors rather than bi-spinors that are the mathematical objects used in the Standard Model to describe fermionic matter. However, the predictive power of tree-level bi-spinor SM for lepto-quark mixing and the existence of supersymmetry that is more compact then the standard one leads us to conjecture that, if the bi-spinor modification of the SM can be constructed and proven to satisfy all precision EW constraints, then bi-spinors could provide a more fitting description of quantum fermionic matter then Dirac spinors.

## Appendix A: Discussion and Perspectives

Having described new realization of supersymmetry, we are naturally led to ask what it all means for building phenomenological models beyond the SM or string models that involve bi-spinors to represent fermionic matter? There are a number of issues that need to be solved before bi-spinor SM can have the same status as the SM and certainly before its s-supersymmetric version can be considered. First, quantum field theory of bi-spinor fields and the corresponding perturbation theory must be defined. This work has been essentially completed [28]. What emerges is that bi-spinor gauge theory admits very specific explicit mass terms and acquires an additional quantum number, the origin of which can be traced to the bi-spinor transformation law of bi-spinors. Second, all observed elementary particles must be classified according to the new quantum number. It turns out that this assignment is essentially unique, if one considers the experimentally observed CKM and PMNS matrix textures [27].

Although a bi-spinor gauge theory is renormalizable by power count, it has dimensionless coupling constants, the formal proof of renormalizability is lacking. At the same time one has to prove that all the electroweak precision measurement constraints that are well but not perfectly satisfied by the SM are satisfied at least as well in bi-spinor version of the SM. Only then one can start discussing reasonably rigorously various s-supersymmetric extensions of bi-spinor SM. All in all, although there is encouraging experimental support in favor of bi-SM, the theory is not sufficiently well developed to compete with the SM or its supersymmetric extensions on equal footing.

We will now turn to various concrete somewhat disjointed topics that obviously need to be explored. One interesting fact about the standard Euclidean lattice supersymmetry is that hypercubic lattice that is typically used for realizations of the standard supersymmetry allows for natural appearance of extended supersymmetry, which is needed to construct twisted realizations of the standard supersymmetry, the generators of which can be enumerated by the big diagonals of the basic lattice hypercube. Whether one can introduce extended s-supersymmetry on a smooth manifold is an open question, the answer to which relies on the full analysis of all possible s-supersymmetric algebras and their irreducible representations. Also the exact connection between s-supersymmetry and supersymmetry of twisted Dirac-Kähler fields on the lattice remains to be explored.

We should comment briefly on the long-standing apparent puzzle about representation of fermions by a collection of anticommuting antisymmetric tensors. From our point of view, the puzzle is resolved through the spinbein decomposition of bi-spinors [22]. In a general spinbein gauge a bi-spinor is completely equivalent to a collection of anti-symmetric tensors. Both are coefficients of expansion of a diform in a particular basis. However, for physical spinbein gauges, where spinbeins must be constant so that they can transfer all their dynamical degrees of freedom to algebraic Dirac spinors, one may consider space-time as if it carries spin. Just like constant energy, on $M_4$ this background space-time spin is not detectable, because of constancy of spinbein. Thus, from the point of view of physical (anti)-Dirac spinors contained in bi-spinors, the vacuum state in bi-spinor gauge theories may be considered as a spin state. Combined with spin of the (anti)-Dirac spinors the vacuum state turns spin of otherwise spin one-half states into spin of integer spin states represented by anti-symmetric tensors.

Since a particular $\xi$-gauge with $\xi_p = 2$ is required for realization of s-supersymmetry, we have to ask ourselves whether such symmetry is physically acceptable. The answer is yes, it is physically acceptable. First, recall from the example of chiral symmetry broken by mass terms that partial symmetries of the Lagrangian can have important physical consequences. Also in the quantum Dirac spinor gauge theory, described by quantum gauge-fixed Lagrangian (28, 29), gauge symmetry is a partial symmetry of the Lagrangian. Similarly, in the quantum gauge-fixed Lagrangian in (28, 29) of the bi-spinor gauge theory only a part of the Lagrangian is gauge invariant. Of course, only "gauge independent" consequences of the presence of symmetry broken or not can be physical. This does not prevent, however, the appearance at some intermediate stages of gauge dependent expressions.

For example, it is well-known that in different gauges different desirable features of a gauge theory become more pronounced or less. We can certainly consider unitarity of *S*-matrix as manifestation of symmetry, a consequence of requirement of conservation of probability. However, this symmetry is transparent only in the unitary gauge. The same applies to covariance of a gauge theory. Only in $\xi$-gauges the covariance is manifest but not in the unitary gauge. In the same vein we can consider $\xi = 2$ gauge of bi-spinor gauge theory as the gauge where s-supersymmetry is manifest or alternatively we can consider it as symmetry of a part of the Lagrangian by analogy with broken chiral symmetry. In either case its effects can be judged only on gauge invariant and measurable quantities. For application to solution of the hierarchy problem, one of such quantities could be a sum of gauge-invariant divergent graphs contributing to Higgs mass. In gauges other then $\xi = 2$ it is not manifest, but what matters of course is whether *S*-matrix amplitudes in some way reflect the presence of partial s-supersymmetry. We will consider this question in more detail elsewhere.

Another interesting aspect of s-supersymmetry is that for asymptotically free theories, the theories that are actually of physical interest, an approximate s-supersymmetry, that is s-supersymmetry of the free part of the Lagrangian, should become exact in the ultraviolet limit. Such asymptotically free s-supersymmetric bi-spinor gauge theories would possess the two desirable features of softly broken standard supersymmetry. At low energy s-supersymmetry is broken, while at high energy it is restored, so that one should expect all the benefits of cancellation of divergences that come from the standard supersymmetry, where one has to break supersymmetry softly.

To indicate how to prove that free-field s-supersymmetry is restored in ultraviolet limit, consider conserved s-supersymmetric currents for free-field case derived in [25], where we refer for details. For complex multiplet these are given via Noether's theorem by

$$J_\varepsilon^{\,\mu} = tr\left[\overline{\overline{\Psi}}(A)\vec{\partial}\gamma^\mu \Psi(\Phi)\right], \qquad \partial_\mu J_\varepsilon^{\,\mu} = 0, \tag{67}$$

$$J_{\varepsilon^*}^{\,\mu} = tr\left[\overline{\overline{\Psi}}(\Phi)\gamma^\mu \vec{\partial} \Psi(A)\right], \qquad \partial_\mu J_{\varepsilon^*}^{\,\mu} = 0, \tag{68}$$

where we defined currents through action variation as

$$\delta S = \delta S(\varepsilon = const) + \int d^4x \left(\partial_\mu \varepsilon\, J_\varepsilon^{\,\mu} + J_{\varepsilon^*}^{\,\mu} \partial_\mu \varepsilon^*\right), \tag{69}$$

while in the chiral multiplet case one obtains

$$J^\mu = tr\left[\overline{\overline{\Psi}}(\Phi_L)\hat{\overline{L}}_\Phi\, \gamma^\mu \left(i\vec{\partial}\right)\Psi(A_R) - \overline{\overline{\Psi}}(A_R)\left(-i\vec{\partial}\right)\gamma^\mu \hat{L}_\Phi \Psi(\Phi_L)\right],$$

$$\partial_\mu J^\mu = 0. \tag{70}$$

It is easy to verify directly that the three interacting currents obtained by replacing derivatives in (67, 68, 70) with their gauge-covariant versions are conserved. Indeed, using appropriate equations of motion we obtain for complex multiplet two covariantly conserved currents

$$J_\varepsilon^\mu = tr\left[\overline{\overline{\Psi}}(A)\bar{\partial}_{A_1}\gamma^\mu\Psi(\Phi)\right], \qquad D_\mu J_\varepsilon^\mu = 0, \qquad (71)$$

$$J_{\varepsilon^*}^\mu = tr\left[\overline{\overline{\Psi}}(\Phi)\gamma^\mu\partial_{A_1}\Psi(A)\right], \qquad D_\mu J_{\varepsilon^*}^\mu = 0, \qquad (72)$$

Analogously, with the help of on-shell equation

$$(\partial_{A_1})\hat{L}_\Phi\Psi_\Phi \equiv \Psi(\nabla_{A_1}L_\Phi\Phi_L) = 0 \qquad (73)$$

we obtain conserved current for the chiral multiplet

$$J^\mu = tr\left[\overline{\overline{\Psi}}_A(-\bar{\partial}_{-A_1})\gamma^\mu\hat{L}_\Phi\Psi_\Phi + \overline{\overline{\Psi}}_\Phi\overline{\overline{L}}_\Phi\gamma^\mu(\bar{\partial}_{A_1})\Psi_A\right], \qquad D_\mu J^\mu = 0, \quad (74)$$

where $D_\mu$ are appropriately defined covariant derivatives, $\partial_{A_1} = \gamma^\mu D_\mu$. Conservation of the currents (71, 72, 74) imply the existence of associated symmetries. These very well may be realized non-linearly, but this is not important for our argument.

We will now turn to asymptotically free theories. These are defined as those for which beta-function vanishes in the ultraviolet limit, which means that at high transfer momentum the effective coupling constant tends to zero and elementary particles cease to interact.

This effect can also be stated in terms of asymptotic behavior of Greens functions of the interacting theory. Roughly, for asymptotically free case they can be represented in an expansion in terms of inverse powers of transfer momentum with the first term corresponding to products of free (renormalized) propagators and the remaining terms suppressed by powers of transfer momentum. The existence of the conserved currents (71, 72, 74) induces corresponding Ward-Takahashi/Slavnov-Taylor identities on the correlators of the bi-spinor gauge theory, defined as T-products of various fields

$$\langle 0|T[\psi(x_1)\cdots\bar{\psi}(x_m)A_\mu(y_1)\cdots A_\nu(y_n)]|0\rangle. \qquad (75)$$

Their derivation follows the standard route by considering

$$D_\lambda\langle 0|T[J^\lambda(x)\psi(x_1)\cdots\bar{\psi}(x_m)A_\mu(y_1)\cdots A_\nu(y_n)]|0\rangle, \qquad (76)$$

where $J^\lambda(x)$ is any of the three currents above. For asymptotically free theory the derived identities would also be representable in inverse power series in transfer momentum with leading term being the identities derived for free theory. Thus in the ultraviolet limit for asymptotically free theories one would expect restoration of s-supersymmetry of (renormalized) free-field Lagrangian in the sense described above: the correlators identities of the interacting case would tend to those of the free case.

Complete and rigorous proof of this conjecture is far beyond the scope of this Letter and we will consider this issue elsewhere. We note, however, recent work [23], where the authors explored a scenario where broken supersymmetry is restored in the ultraviolet limit by quantum effects.

## Acknowledgement

I wish to thank the referees of the paper for posing interesting and significant questions that need to be answered before full-fledged bi-spinor SM can be constructed.